\documentstyle[ltwol]{article}

\def\Journal#1#2#3#4{{#1} {\bf #2}, #3 (#4)}

\def\NPB{{\em Nucl. Phys.} B}

\def\PRD{{\em Phys. Rev.} D}  
\def\ZPC{{\em Z. Phys.} C}

\def\ra{\rightarrow}

\def\be{\begin{equation}}
\def\ee{\end{equation}}
\def\bea{\begin{eqnarray}}
\def\eea{\end{eqnarray}}

\input psfig

\begin{document}
\titlepage
\begin{flushright}
IFT 98/12 \\
{\bf hep-ph/9810253}
\end{flushright}
\vskip 3cm
\centerline{\bf {\Large {{NLO PREDICTION FOR THE PHOTOPRODUCTION}}}}
\vskip 0.4cm 
\centerline{\bf {\Large {OF THE ISOLATED PHOTON AT HERA}}}
\vskip 4cm
\centerline{\Large Maria Krawczyk$^*$ and Andrzej Zembrzuski}                 
\centerline{Institute of Theoretical Physics, Warsaw University,}
\centerline{ul. Ho\.za 69, 00-681 Warsaw, Poland}
\centerline{E-mail: krawczyk@fuw.edu.pl,azem@fuw.edu.pl}

\vskip 4cm
\begin{abstract}
The NLO calculation of the DIC process with the isolated photon
is presented and the
comparison is made \newline
with the other NLO  calculation and
with recent 
measurement at HERA. The satisfactory agreement was found.
\end{abstract}

\footnotetext{$^*$ Talk at ICHEP'98, Vancouver, Canada, 23-29 July 1998} 

\title{
NLO PREDICTION FOR THE PHOTOPRODUCTION OF THE ISOLATED PHOTON AT HERA
}

\author{M. KRAWCZYK, A. ZEMBRZUSKI}

\address{Institute of Theoretical Physics, Warsaw University,
ul. Ho\.za 69, 00-681 Warsaw, 
Poland\\E-mail: krawczyk@fuw.edu.pl,azem@fuw.edu.pl}


\twocolumn[\maketitle\abstracts{ 
The NLO calculation of the DIC process with the isolated photon
is presented and the
comparison is made with the other NLO  calculation and with recent 
measurement at HERA. The satisfactory agreement was found.
}]

\section{Introduction}
We consider Deep Inelastic Compton (DIC)  scattering where the photon
with large transverse momentum $p_T$ is produced 
 in $ep$ collision. 
With antitagging or untagging conditions 
such reaction is  dominated by events with 
almost real photons mediating the $ep$ interaction, 
$Q^2\approx 0$, so in practice the photoproduction is considered.
 The observed final $\gamma$
 arises directly from the direct subprocess $\gamma q\ra \gamma q$
or from subprocesses where partonic content of 
the photon contributes.
The final photon may originate also from the fragmentation processes, 
where  $q$ or  $g$ 'decays' into $\gamma$.

In  this talk the results of the 
NLO calculation for DIC process \cite{mkzz} are shown
with the discussion on the choice of the relevant set of diagrams.
Then the influence of the
isolation cuts on the production rate of the final photon
is shown and the role of additional cuts applied in
the experimental analysis of this process at HERA is discussed.
The importance of the box diagram $\gamma g\ra \gamma g$,
being the higher order direct process, is stressed.

 The final results for the isolated photon production in the DIC process  at 
HERA are compared
with the existing in literature calculation \cite{gor98}  and with the corresponding
data \cite{zeus}.

\section{Deep inelastic Compton process in NLO}
\subsection{NLO calculation for processes involving photons}\label{subsec:prod}
We start  by describing processes which are (should be?)
included in the NLO QCD calculation
of the cross section for  the DIC process, 
\be
\gamma p \ra \gamma X.
\ee
Although we will discuss the process (1), the problem which we
touch upon   is more general -  it is related to the different approaches
to   
NLO calculations of  cross sections for hadronic  processes involving photons.
\begin{figure}
\center
\vskip 0.5cm
\psfig{figure=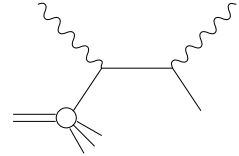,height=1.5cm}
\caption{The lowest order Born contribution}
\label{fig:born}
\end{figure}

\begin{figure}
\center
\vskip 0.5cm
\psfig{figure=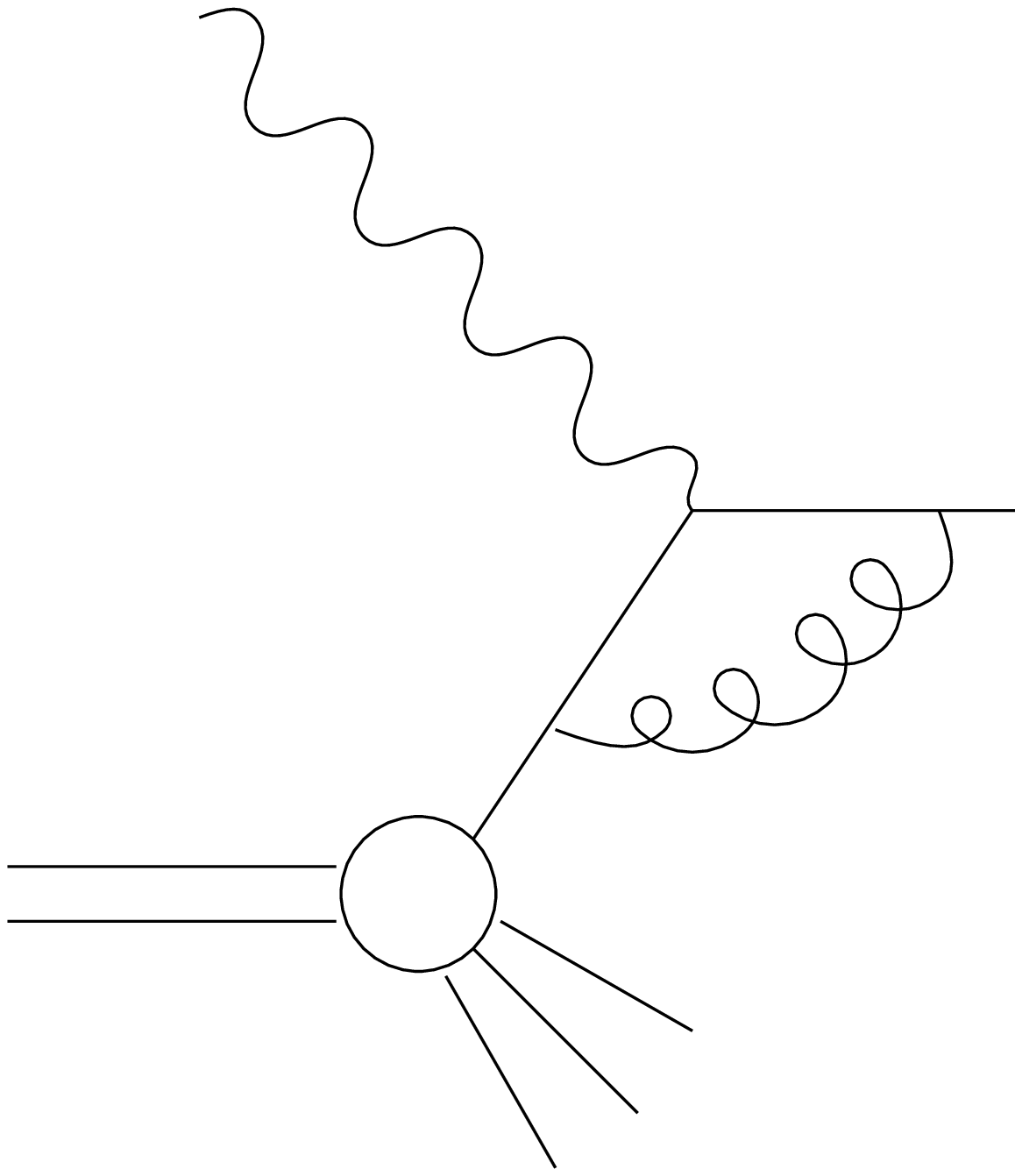,height=1.5cm}
\vskip -1.5cm
\psfig{figure=real.ps,height=1.5cm}
\caption{Examples of the virtual and real $\alpha_s$ corrections
to the Born contribution}
\label{fig:cor}
\end{figure}

The Born level contribution to the cross section, the lowest order in the 
strong  coupling constant $\alpha_s$ term,  is based on the Compton process 
on the quark (Fig.~\ref{fig:born}):
\be
\gamma q\ra \gamma q.
\ee
It leads to  the  [$\alpha_{em}^2$] order contributions to
 the  partonic cross section, and at the same $\alpha_{em}^2$  order
it contributes to  the hadronic cross section 
for the process  $\gamma p\ra \gamma X$. In the NLO analysis
the $\alpha_s$ corrections to (2)
are calculated and the terms of order  $\alpha_{em}^2
\alpha_s$ appear (Fig.~\ref{fig:cor}).
 
\begin{figure}
\center
\vskip 1cm
\psfig{figure=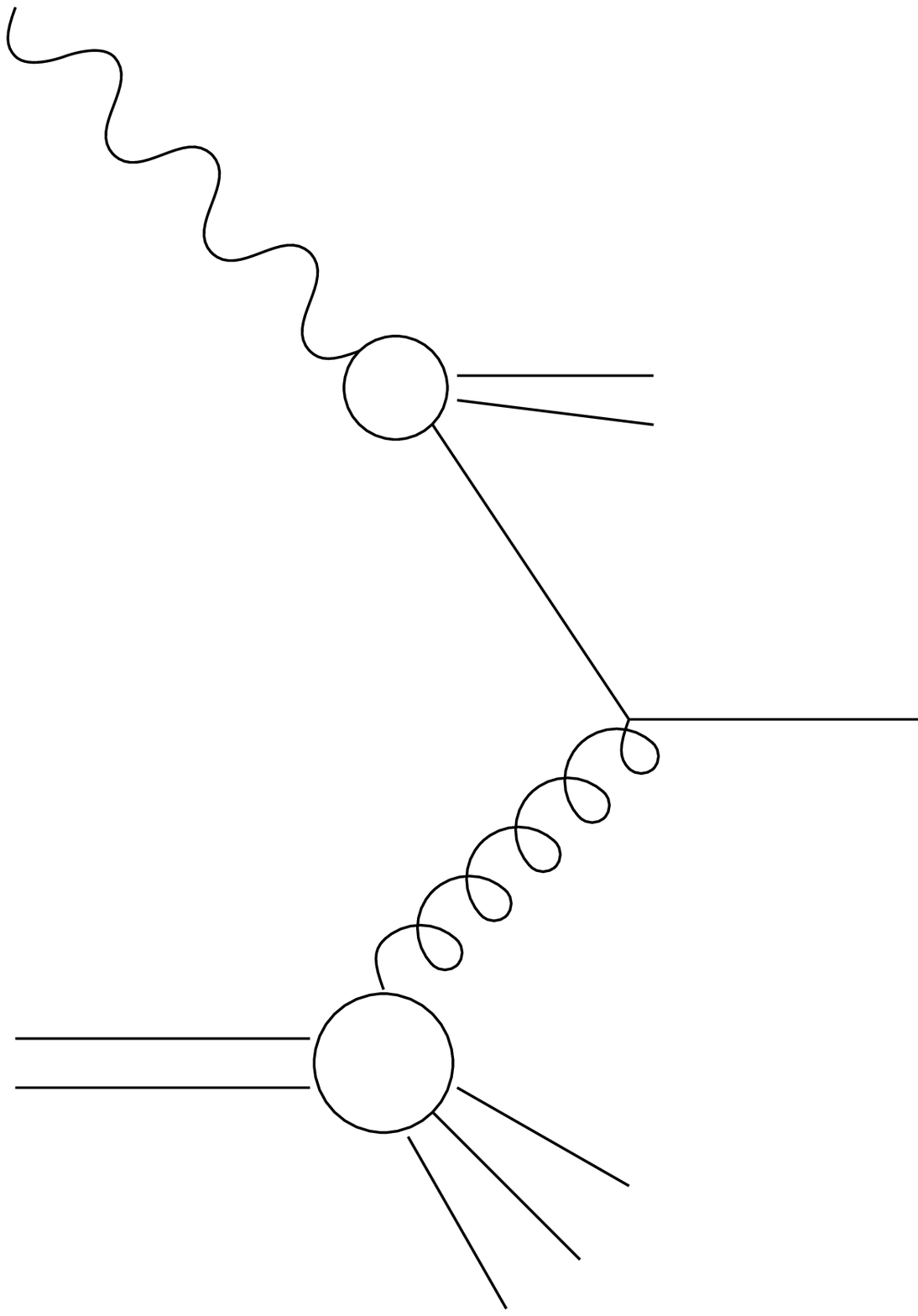,height=1.5cm}
\vskip -1.5cm
\psfig{figure=frag.ps,height=1.5cm}
\caption{Examples of single resolved processes.
a) the resolved initial photon, b) the resolved final photon}
\label{fig:1res}
\end{figure}

\begin{figure}
\center
\vskip 1cm
\psfig{figure=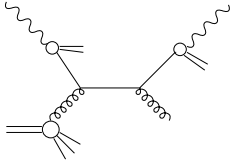,height=1.5cm}
\caption{An example of double resolved photon process}
\label{fig:2res}
\end{figure}

The  Parton Model prediction for the DIC process,
 which applies for $x_T=2p_T/\sqrt{S}\sim{\cal O}(1)$,
relies solely on the Born contribution (2) \cite{dic-pm}.
  For semihard processes, where $x_T \ll 1$, the  prediction
based on (2) is
 not a good approximation, since now 
 one should consider besides (2)
  also contributions involving interactions of the partonic
content of the photon(s). There are two classes of such contributions:
{\underline  {single resolved}} with resolved initial $or$ 
final photon, and {\underline{ double resolved}} with
both the initial $and$ the final photon resolved
(Figs.~\ref{fig:1res}, \ref{fig:2res}).
They give contributions to partonic cross sections of orders 
[$\alpha_{em} \alpha_s$] (single resolved) and [$\alpha_s^2$] 
(double resolved).
 If one takes into account that partonic densities in the photon and
the parton fragmentation into the photon are of order $\sim \alpha_{em}$, 
the final contribution to the hadronic cross section from resolved 
photon processes are  $\sim \alpha_{em}^2 \alpha_s$ and 
$\alpha_{em}^2 \alpha_s^2$, respectively. 
{In the LLO approach 
these three types of subprocesses:  with two direct photons, with
one and two resolved photons,
  convoluted with the relevant LL parton densities, are considered
\cite{dic-ll}.}

 In the NLO approach,
among  $\alpha_{em}^2 \alpha_s$  terms,
 there are terms related to 
the  collinear singularities, 
which then have to be substracted and treated as 
a part of the corresponding {\underline {single}}
 resolved photon contribution. 
At the same time in the NLO expression there are no 
 collinear singularities
 terms which would correspond to the {\underline {double}}
 resolved photon contributions.
It  indicates that taking into account subprocesses
[$\alpha_s^2$] associated with the both initial and final photons resolved
goes  beyond the accuracy of the NLO calculation.
This will be consistent
 within the  NNLO approach, where $\alpha_s^2$ correction to the Born term and
$\alpha_s$ correction to the single resolved terms should be included,
all giving the same  $\alpha_{em}^2\alpha_s^2$ order contribution to the
 hadronic cross sections.

The other set of diagrams is considered by some authors
in the NLO approach to DIC process (1), due to 
 their different 
 way of the  counting the order of  the parton densities in the  photon 
(or the parton fragmentation into the photon) \cite{dic-nlo}. 
This approach, which we will call
``${{1}\over{\alpha_s}}$'' approach to the structure of photon,
is motivated by  the existence of the 
large logarithms of $Q^2$ in the $F_2^{\gamma}$ already in PM.
By expressing  $\ln (Q^2/\Lambda_{QCD}^2)$ as $\sim 
{{1}\over{\alpha_s}}$ 
one  treats
the parton densities in photon as proportional to  $\alpha_{em}/\alpha_s$.
By applying this method to the DIC process, we see that here
the single resolved photon contribution to the hadronic cross section
is of the same order as the Born term, namely 
\be
{{\alpha_{em}}\over{\alpha_s}}\otimes [  \alpha_{em} \alpha_s]  
\otimes 1=\alpha_{em}^2,
\ee
also the
double resolved photon contribution is of the same order 
\be
{{\alpha_{em}}\over{\alpha_s}}\otimes [ \alpha_s^2]  \otimes 
 {{{\alpha_{em}}}\over{\alpha_s}}=
\alpha_{em}^2.
\ee
We see that this way,
 the same $\alpha_{em}^2$ order contributions
to the hadronic cross section are given by the Born direct process,
 single resolved photon  and double resolved photon processes,
although  they correspond to    quite different final states 
(observe a lack of the remnant of the photon in the direct process).
Moreover, they constitute the lowest order
(in the strong coupling constant) term in the perturbative expansion,
actually the zeroth order, so 
 the (leading) dependence on the strong coupling constant is absent.  
If  one takes into account that some of these terms  correspond
 to 
 the hard processes involving gluons,  a lack of $\alpha_s$ coupling 
in the cross section   seems to be contrary to the intuition.
On the other hand, on such set of contributions the LL prediction is based.

In this approach beside  the $\alpha_s$ correction to
 the Born process
also the  $\alpha_s$ corrections to the  single  and to the double
resolved photon
processes should be included in the NLO calculation,
since all of them give terms of the same order, $\alpha_{em}^2{ \alpha_s}$.

To summarize, the first NLO approach starts with one basic, direct  
subprocess,
while the second one with three different
 subprocesses (as in LO). 
Obviously, some of NNLO terms in the first method 
belong to the NLO terms in the second one.
\subsection {$\alpha_{em}^2$, \protect{$\alpha_{em} \alpha_{s}$} 
and $\alpha_s^2$  subprocesses in DIC}
 Below we will discuss our NLO analysis of the DIC process \cite{mkzz},
where the parton densities in the photon (the parton fragmentation 
into the photon) are treated as $\sim$ $\alpha_{em}$.

In the NLO calculation of the DIC process we take the following subprocesses
into account:\\
$\bullet$   the  Born contribution (2) (Fig.~\ref{fig:born})\\
$\bullet$  the $\alpha_s$ corrections to the Born diagram
(Figs.~\ref{fig:cor})\\
$\bullet$ two types of single resolved photon contributions, 
with the resolved initial and 
with the resolved final photon
(Figs.~\ref{fig:1res})

\begin{figure}
\center
\vskip 1.5cm
\psfig{figure=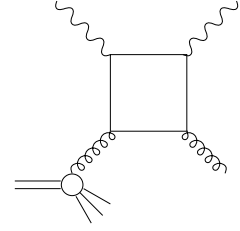,height=1.5cm}
\caption{The box diagram}
\label{fig:box}
\end{figure}

 Beside the above   set of diagrams, the full NLO set, 
  we will in addition  include two  terms
 of order $\alpha_{em}^2 \alpha_s^2$ (formally  from the NNLO set):
 the double resolved contributions 
(Fig.~\ref{fig:2res})
and the direct box diagram $\gamma g \ra \gamma g$ (Fig.~\ref{fig:box}),
since they were found to be large \cite{mk}. 
\subsection{Inclusive cross section for the DIC process at $ep$ colliders} 
Infrared singularities which appear in the NLO calculations
of the virtual corrections to the
Born process are canceled by infrared singularities in the
real corrections. 
 All collinear (mass) singularities
are factorized into structure and fragmentation functions,
as discussed above, leading to the final formula for the
 inclusive cross section for the $\gamma p\rightarrow\gamma X$
scattering:
\begin{eqnarray}
E_{\gamma}{d^3\sigma^{\gamma p\rightarrow\gamma X}\over
d^3p_{\gamma}} = ~~~~~~~~~~~~~~~~~~~~~~~~~~~~~~~~~~~~~~~~~~~~~~~~~~~\\ \nonumber
\sum\int {dz\over z^2}\int dx_{\gamma}\int dx
f_{a/\gamma}(x_{\gamma},\bar{Q}^2) f_{b/p}(x,\bar{Q}^2)\cdot \\ \nonumber
\cdot D_{\gamma /c}(z,\bar{Q}^2)
E_{\gamma}{d^3\sigma^{ab\rightarrow cd}\over d^3p_{\gamma}}
+\sum\int dx f_{b/p}(x,\bar{Q}^2) {\alpha_S\over 2\pi^2 \hat{s}}
K_b,
\end{eqnarray}
where the second term, K-term, describes the QCD corrections to the
Born process and the first term is a sum over all the other
contributions. The  $f_{b/p}$ is a $b$ parton distribution in
the proton while $f_{a/\gamma}$ is a $a$ parton
distribution in the photon. For direct initial photon 
(where $a=\gamma$):
$f_{a/\gamma}=\delta (x_{\gamma}-1)$.
$D_{\gamma /c}$ is a $c$ parton fragmentation function
into a photon, or for non-fragmentation processes
(where $c=\gamma$) $D_{\gamma /c}=\delta(z-1)$.
$x$, $x_{\gamma}$, $z$ are: a part of the initial proton
momentum taken by the $b$ parton, a part of the initial
photon momentum taken by the $a$ parton, and a part of
the $c$ parton momentum taken by the final photon, 
respectively. The $\bar Q$ scale is provided here by the $p_T$. 

Denoting the differential cross section for the $\gamma p$ scattering (1) 
by $\tilde {\sigma}^{\gamma p}$ 
we now relate it to the corresponding $ep$ cross section.
The differential cross section for the $ep$ scattering  
in the antitagging conditions  can be calculated using 
the equivalent photon approximation:
\begin{eqnarray}
\tilde {\sigma}^{ep}=\int G_{\gamma/e}(y) 
\tilde {\sigma}^{\gamma p} dy ,
\end{eqnarray}
where  $y=E_{\gamma}/E_e$ is
a fraction of the initial electron energy taken by the
photon, and the (real) photon 
distribution in the electron is given by:
\begin{eqnarray}
G_{\gamma/e}(y)={\alpha\over 2\pi}\cdot
~~~~~~~~~~~~~~~~~~~~~~~~~~~~~~~~~~~~~~~~~~~~~~~ \\ \nonumber
\cdot\{ {1+(1-y)^2\over y}
\ln [{Q^2_{max}(1-y)\over m_e^2 y^2}]\!-\!2(1-y)
+{2m_e^2y^2\over Q^2_{max}}\},
\end{eqnarray}
with $m_e$ being the electron mass. The $Q^2_{max}$ value
we assume as equal to 1 GeV$^2$, which is typical for the 
photoproduction measurements at HERA collider.
\subsection{Isolated photon production in DIC process }\label{subsec:wpp}
In order to reduce backgrounds from $\pi^0$'s and $\gamma$'s
radiated from final state hadrons isolation cuts on the
observed photon have
to be performed. Experimentally  the isolation cuts are 
defined by demanding that  a sum of 
hadronic energy within a cone of radius $R$ around the final 
photon should be smaller then the final photon energy multiplied
by parameter $\epsilon\sim 0.1$:
\begin{eqnarray}
\sum_{hadrons}E_h<\epsilon E_{\gamma}.
\end{eqnarray} 
The radius is defined in the rapidity and azimutal
angle phase space: $R=\sqrt{\Delta\Phi^2+\Delta\eta^2}$.

The simplest way to calculate the differential cross section
for isolated photon, $\tilde{\sigma}^{isol}$, is to calculate
difference of an inclusive differential cross section,
$\tilde{\sigma}^{incl}$, and a subtraction term,
$\tilde{\sigma}^{subt}$ (see \cite{vog}):
\begin{eqnarray}
\tilde{\sigma}^{isol}=\tilde{\sigma}^{incl}-\tilde{\sigma}^{subt}.
\end{eqnarray}
By the subtraction term we mean differential cross section
with cuts opposite to the isolation cuts, i.e.
within a cone of radius $R$ around the final photon
 there should appear hadrons with total energy 
higher than the photon energy multiplied by $\epsilon$:
\begin{eqnarray}
\sum_{hadrons}E_h>\epsilon E_{\gamma}.
\end{eqnarray} 
The cuts are imposed in the partonic phase space integration (for 
K-term) and integration over $z$ (for fragmentation processes).

We calculate the subtraction term in an approximate way,
with two simplifying assumtions \cite{azem}:\newline 
$\bullet$ the parameter $\epsilon$ is small, 
$\epsilon\ll 1$\newline
$\bullet$ an angle $\delta$ between the final photon and 
a parton (from which a hadronic jet arise) inside the cone 
is small (i.e. the cone is small)
and can be approximated by
$\delta=R/cosh(\eta)$, where $\eta$ is the rapidity of the
photon.

The above approximations are used only when calculating the 
K-term (i.e. the QCD corrections to the Born process)
in the subtraction term $\tilde{\sigma}^{subt}$.
Because this term gives about 4\% of the
inclusive cross section, we expect that the error
resulting from using the approximations is negligible.
\begin{figure*}
\center
\vskip 1.5cm
\psfig{figure=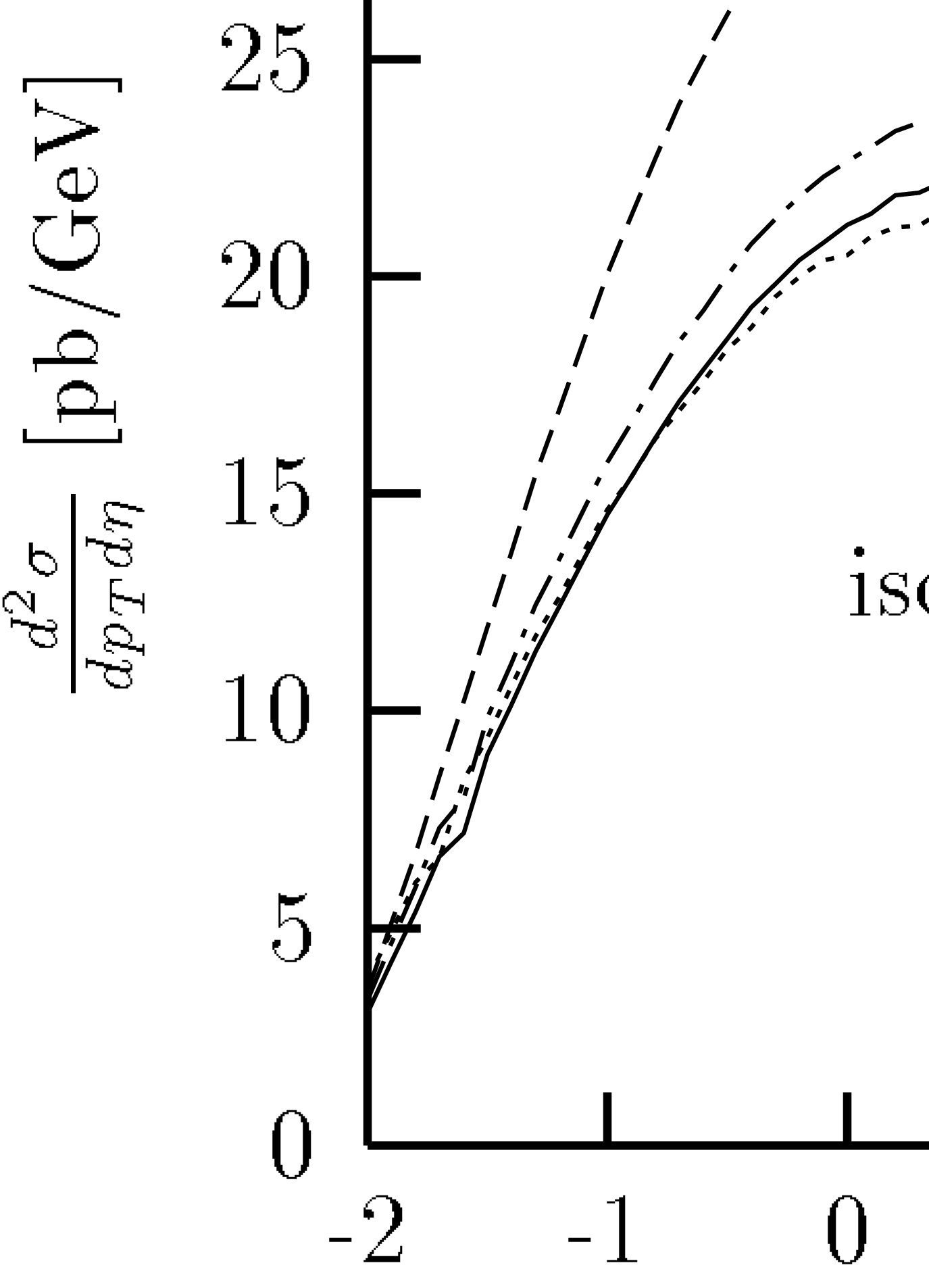,height=3cm}
\vskip -3cm
\psfig{figure=fig3.ps,height=3cm}
\vskip -3cm
\psfig{figure=fig5.ps,height=3cm}
\caption{The photon rapidity distributions. a) The differential cross section
$d^2\sigma /dp_Td\eta$ integrated over whole range of $y$ for $p_T$=5 GeV 
for the final photon  not isolated (upper line) or isolated with R=1 and 
$ \epsilon$=0.05, 0.1, 0.2.
b) The cross section $d\sigma /d\eta$ integrated over 0.16$ \le y \le$0.8 and
5 GeV$ \le p_T\le $10 GeV for the isolated final 
photon with R=1 and $ \epsilon$=0.1. 
 The result for $N_f$=4 (solid line) and for 
comparison the same without the box contribution
(dotted line). The result for $N_f$=3 (dashed line) is also shown.
c) The cross section as in b) for $N_f$=4 for 
the isolated photon plus jet production, where the jet obeys 
conditions: -1.5$ \le \eta^{jet} \le $1.8 and  $p_T^{jet} \ge $5 GeV
(solid line).The data are the ZEUS experiment preliminary
results  $^{3}$.}
\label{fig:fig235}
\end{figure*}
\section{Results}
\subsection{Inclusive cross section}
First we will discuss the general features of the inclusive 
cross section for the DIC process in the $ep$ collision.
We take the HERA collider energies: $E_e$=27.5 GeV, $E_p$=820 GeV 
and we limit ourselves to the $p_T$ range of the final photon between 
5 and 10 GeV ($x_T$ is 0.03-0.07).
 The calculation was performed in ${ \overline {\rm MS}}$ scheme and 
 as a  hard (renormalization, factorization) scale we take
$\bar{Q}=p_T$, also $\bar{Q}=p_T/2$ and $2p_T$ was studied. 
Number of active flavours is assumed to be $N_f$=4 (and for comparison
 $N_f$=3).
$N_f$ enters into the evaluation of the 
coupling constant $\alpha_s$.
Note that the assumed NLO form of $\alpha_s$ plays a role, as
 changing the form from: 
\be
\alpha_s={{4 \pi}\over {\beta_0 \ln(\bar{Q}^2/\Lambda_{QCD}^2 )}}
{{1}\over{1+{{2\beta_1}\over{\beta_0^2}} 
{{\ln[\ln(\bar{Q}^2/\Lambda_{QCD}^2)]}\over{\ln(\bar{Q}^2/\Lambda_{QCD}^2)}}}}
\ee
to
\be
\alpha_s={{4 \pi}\over {\beta_0 \ln(\bar{Q}^2/\Lambda_{QCD}^2) }}
[1-{{2\beta_1}\over{\beta_0^2}} {{\ln[\ln(\bar{Q}^2/\Lambda_{QCD}^2)]}
\over{\ln(\bar{Q}^2/\Lambda_{QCD}^2)}}]
\ee
($\beta_0=11-2/3 N_f$ and $\beta_1=51-19/3 N_f$) changes the results by 5\%.
The final results will be given using the latter form with
 parameters: $\Lambda_{QCD}$=0.2 GeV for $N_f$=4 
and 0.248 GeV for  $N_f$=3.

We use Gl\"uck, Reya, Vogt NLO parametrizations of the proton
structure function, the photon structure function
 and the fragmentation function \cite{grv}, for comparison
 also other photon structure functions are used \cite{acfgp,gs}.
The value $\Lambda_{QCD}$ was used here as fitted to the data.

The importance of the particular contributions 
to the cross section can be illustrated by 
the following results obtained 
for the {\sl inclusive} cross section integrated over 5 GeV$<p_T<$10 GeV
(where the sum over all subprocesses is equal to 210 pb):
$Born=41.2\% $, $single$ $resolved=34.2\% $,
$double$ $resolved=15.7\%$, $box=5.4\% $, $K$-$term$=3.9\%.
So it is obvious that the single resolved photon processes
give contribution comparable to the Born term. Also the double resolved
 photon processes are  important giving half of the single 
resolved photon process contribution. The overall double resolved photon
cross section is build from   many, relatively small, individual terms.
The box diagram gives 14\% of the Born contribution-
relatively large box contribution is due to  large 
 gluonic content of the proton  at small $x_p$ (the Born process
bases on the quark density at the same value $x_p$). 
\subsection{Inclusive versus isolated photon cross section} 
The isolation cut was introduced with
the radius $R=1$, and $\epsilon$=0.05, 0.1 and 0.2 .
 Results are presented in Fig.6a
where the {\it inclusive} cross section and the {\it isolated} photon
 cross sections are compared.
Large effect is seen when the isolation cut is imposed on the
final photon  in DIC process, with a larger suppression of the fragmentation
processes (up to 80\%, see \cite{azem}) as expected.
This effect is not very sensitive to the value of $ \epsilon $.
\subsection{Comparison with data}
To compare our results with data 
the calculations were performed for the HERA collider energies
with cuts used in  the ZEUS experiment \cite{zeus}.
For the isolated photon production in DIC process 
we take  $R=$1, $\epsilon=$0.1 and consider 
ranges: $~~~5~ {\rm GeV}\le p_T\le 10 ~{\rm GeV}~,~
0.16\le y\le  0.8.$\\
Two types of final state were measured:\\
''$\gamma$''-- an isolated photon with rapidity $-0.7\le\eta^{\gamma}\le0.9$,\\
``$\gamma$+jet''-- the isolated photon as above plus 
the jet with \\
~~~~~~~~~~$-1.5\le\eta^{jet}\le 1.8~,~~~ p_T^{jet}\ge 5 ~~{\rm GeV}.$

In Fig.6b the  isolated photon rapidity distributions with the above criteria
are shown. The importance of the box diagram is seen (the double resolved
contribution is smaller than box up to  factor 2 for negative
 $\eta^{\gamma}$).
Note also a large difference between results for $N_f$=4 and 3
(due to fourth power of electric charge
characterizing  processes involving two photons).
The prediction based on the four  massless quarks overestimates 
the production rate, while this with $N_f$=3 underestimates this rate.
We belive that the true prediction lies between two curves, but
closer to the upper one.
(The proper treatment of the charm quark is crucial here
and will be discussed  elsewhere \cite{kto} ).
In Fig.6c the results for the final state with a photon 
and the jet are presented.
The calculation 
bases here on the additional assumption and therefore
should be treated as an estimation of the considered cross section
(see \cite{azem} for details ).
Resonable agreement with data are found for both final states (Figs.6b,c).

Below we discuss results for the isolated photon cross sections  
within the ZEUS cuts listed above, and integrated over 
0.8$\le x_{\gamma}\le 1$ -- a ``direct'' sample.
Particular
 contributions to 
the isolated photon cross section  
are   presented in Table 1 for ``$\gamma$'' and ``$\gamma$+jet'' 
in the final state. (Results correspond to  {$N_f$=4}).
The  total contribution for ``$\gamma$+jet''  is smaller by 2 pb 
as  compared to ``$\gamma$'', nevertheless
 the following ratios have  similar
pattern in both cases: ``Born/tot''=65\% and ``box/tot''=13\%.
We see how much the role of the direct processes is enhanced
by the isolation cut and the range of $\eta^{\gamma}$. Note also that now
 ``box/Born''=20\%, while the ``2res/Born'' $\sim$2\%.
 
Results  obtained using  other choices of the scale $\bar Q$:
$\bar Q$= $p_T/2$, $2p_T$, differ from ones based on 
the reference scale $p_T$ by $\pm$ 4\% (6\%) for ``$\gamma$''
(``$\gamma$+jet'') final state. 
The results based on ACFGP or GS
parton parametrization for the photon differ from GRV predictions
  by less than 4\%.

   The ZEUS measurement for the  ``$\gamma$+jet'' final state leads to 
cross section equals to $15.4\pm1.6\pm2.2$ pb \cite{zeus}.
\begin{table}
\begin{center}
\caption{DIC cross section (pb) for 0.8$ \le x_{\gamma}\le$1}\label{tab:smtab}
\vspace{0.2cm}
\begin{tabular}{|c|c|c|c|c|} 
\hline 
\raisebox{0pt}[12pt][6pt]{$subprocess$} & 
\raisebox{0pt}[12pt][6pt]{$tot$} & 
\raisebox{0pt}[12pt][6pt]{$Born$} &
\raisebox{0pt}[12pt][6pt]{$box$} &
\raisebox{0pt}[12pt][6pt]{$2res$} \\
\hline
\raisebox{0pt}[12pt][6pt]{$\gamma$} & 
\raisebox{0pt}[12pt][6pt]{25.2} & 
\raisebox{0pt}[12pt][6pt]{16.3} & 
\raisebox{0pt}[12pt][6pt]{3.2} &
\raisebox{0pt}[12pt][6pt]{0.39}\\
\hline
\raisebox{0pt}[12pt][6pt]{$\gamma+jet$} & 
\raisebox{0pt}[12pt][6pt]{23.4} & 
\raisebox{0pt}[12pt][6pt]{15.1} & 
\raisebox{0pt}[12pt][6pt]{3.0} &
\raisebox{0pt}[12pt][6pt]{0.27} \\ 
\hline
\end{tabular}
\end{center}
\end{table}
\vspace*{3pt}
\subsection{Comparison with other results}
Our NLO calculation of DIC process 
differs from NLO analysis
presented in Ref.[2],
based on the ``$1/\alpha_s$'' approach,
by types of subprocesses included.
In our analysis we do not include $ \alpha_s $ 
corrections to the single resolved processes, the NNLO term. 
On the other hand we include the box diagram,  
although being also beyond the NLO accuracy, it  was found to be 
large especially in the experimental conditions used in the ZEUS experiment.
Our prediction  (GRV parton parametrizations with
 $N_f=4$, scale $\bar Q$=$p_T$) for the rapidity distribution for 
the  ``$\gamma$''
  is lower by 10\% to 25\% 
for the negative to positive rapidity than in Ref.[2], while
for the ``$\gamma$ +jet'' final state
our results are higher by 25\% to 10\%, respectively.
\section{Conclusion} 
Results of the NLO calulation for the isolated $ \gamma$ 
production in DIC process at HERA are presented. 
In addition two NNLO contributions: 
double resolved photon processes and box diagram were studied.
The satisfactory agreement with data is obtained for both
``$\gamma$'' and ``$\gamma$+jet'' events for the rapidity distribution.
The box diagram contributes 13\% to the ``direct'' sample
as measured by ZEUS group.

The agreement with the other NLO calculation of the DIC, 
which based on a different set of diagrams, is obtained within 10 to 25\%. 
\section*{Acknowledgements}
We thanks P. Bussey and L. Gordon for important discussions.
MK is grateful to Stan Brodsky for useful discussions and 
hospitality during her stay at SLAC.

Supported by Grant No 2P03B18410.

\end{document}